\documentclass[conference]{IEEEtran}
\IEEEoverridecommandlockouts
\usepackage{amsmath,amssymb,amsfonts}
\usepackage{algorithmic}
\usepackage{subcaption,graphicx}
\usepackage{graphics}
\usepackage{wrapfig}
\usepackage{textcomp}
\usepackage{color}
\usepackage{soul}
\usepackage{booktabs}
\usepackage{orcidlink}
\usepackage{nomencl}
\makenomenclature

\newcommand{\bS}{\mbox{\boldmath $S$}}

\newcommand{\bV}{\mbox{\boldmath $V$}}
\newcommand{\bW}{\mbox{\boldmath $W$}}

\newcommand{\Real}{\mathbb R}

\newcommand{\be}{\begin{eqnarray}}
\newcommand{\ee}{\end{eqnarray}}
\newcommand{\orcid}[1]{\href{https://orcid.org/#1}{\textcolor[HTML]{A6CE39}{\aiOrcid}}}

\hypersetup{
colorlinks=true,
linkcolor=red,
urlcolor=green,
citecolor=blue
}

\begin{document}

\title{Non-negative matrix underapproximation as optimal frequency band selector }

\author{\IEEEauthorblockN{1\textsuperscript{st}Mateusz Gabor \orcidlink{0000-0002-5397-0655}}
\IEEEauthorblockA{\textit{Faculty of Electronics, Photonics, and Microsystems} \\
\textit{Wroclaw University of Science and Technology}\\
Wybrzeze Wyspianskiego 27, 50-370 Wroclaw, Poland \\
mateusz.gabor@pwr.edu.pl}
\and
\IEEEauthorblockN{2\textsuperscript{nd} Rafał Zdunek \orcidlink{0000-0003-3323-6717} }
\IEEEauthorblockA{\textit{Faculty of Electronics, Photonics, and Microsystems} \\
\textit{Wroclaw University of Science and Technology}\\
Wybrzeze Wyspianskiego 27, 50-370 Wroclaw, Poland  \\
rafal.zdunek@pwr.edu.pl}
\and
\IEEEauthorblockN{3\textsuperscript{rd} Radosław Zimroz \orcidlink{0000-0003-4781-9972}}
\IEEEauthorblockA{\textit{Faculty of Geoengineering, Mining and Geology} \\
\textit{ Wroclaw University of Science and Technology}\\
Na Grobli 15, 50-421 Wroclaw, Poland \\
radoslaw.zimroz@pwr.edu.pl}
\and
\IEEEauthorblockN{4\textsuperscript{th} Agnieszka Wyłomańska \orcidlink{0000-0001-9750-1351}}
\IEEEauthorblockA{\textit{Faculty of Pure and Applied Mathematics, Hugo Steinhaus Centre} \\
\textit{Wroclaw University of Science and Technology}\\
Janiszewskiego 14a, 50-372 Wroclaw, Poland \\
agnieszka.wylomanska@pwr.edu.pl}
}

\maketitle

\begin{abstract}
Time-frequency representation (TFR) is often used for non-stationary signal analysis. The most intuitive and interpretable TFR is the spectrogram. Recently, a concept of non-negative matrix factorization (NMF) has been successfully applied to local damage detection in rolling elements of bearings via spectrogram factorization. NMF applied to the spectrogram allows one to find an informative frequency band, which could be further used as a filter characteristic. However, the obtained filter characteristics mostly detect the informative frequency band, which also encompasses a lot of noise. In the case where noise is more problematic, as is the case for acoustic signals from industrial machines, the NMF hardly detects the damage. To solve this problem and obtain more selective filters, which are more robust to noise, we propose the non-negative matrix under-approximation (NMU) as an informative frequency band selector. Due to the more sparse parts-based representation of the NMU compared to NMF, NMU provides more selective filter characteristics, which neglect the non-informative frequency bands related to the noise. In practice, it means that NMU gives a better signal-to-noise ratio for the filtered signal. The efficiency of the proposed approach has been validated on the vibration signal from the test rig and the acoustic signal from an idler.
\end{abstract}
\begin{IEEEkeywords}
fault detection, bearings, non-negative matrix under-approximation, spectrogram, Gaussian noise
\end{IEEEkeywords}

\nomenclature{TFR}{Time-frequency representation}
\nomenclature{NMF}{Non-negative matrix factorization}
\nomenclature{NTF}{Non-negative tensor factorization}
\nomenclature{NMU}{Non-negative matrix underapproximation}
\nomenclature{SOI}{Signal of interest}
\nomenclature{EMD}{Empirical mode decomposition}
\nomenclature{OSF}{Order statistics filter}
\nomenclature{IFB}{Informative frequency band}
\nomenclature{OFB}{Optimal frequency band}
\nomenclature{STFT}{Short-time Fourier transform}
\nomenclature{G-NMU}{Global non-negative matrix underapproximation}
\nomenclature{R-NMU}{Recursive non-negative matrix underapproximation}
\nomenclature{DFT}{Discrete Fourier transform}
\nomenclature{SK}{Spectral kurtosis}
\nomenclature{SVD}{Singular value decomposition}

\printnomenclature

\section{Introduction}
\label{sec:introduction}
One of the challenges in diagnostics is local damage detection in bearings, where the impulsive periodic SOI is hidden under heavy noise in the measured signal. The detection is mostly done on the basis of the impulsiveness or periodicity of the SOI, or a combination of both. 

In most practical cases, the raw vibration signal has a complicated spectral structure. An intuitive approach is to decompose the observed signal and then to process sub-signals that have a simpler structure. Recently, Cheng et al. proposed the method for the identification of informative sub-signals (sub-bands) and their integration for the combined square envelope spectrum analysis \cite{Cheng20232495}. Li et al. \cite{Li2023567} proposed a fast and adaptive EMD that combines the advantages of the OSF with the original EMD. Another method is the automated variational non-linear chirp mode decomposition for bearing fault diagnosis, which was proposed by Dubey et al. \cite{Dubey20231}. Li {\it et al.} \cite{Li20213220} proposed the combination of enhanced SVD with the wavelet packet transform to select an appropriate informative frequency band. The sparse Bayesian approach was proposed by Cao {\it et al.} in \cite{cao2023sparse}.

The TFR could also be considered as a signal decomposition because we can track the variation of energy in time at a given frequency band. There are many types of time-frequency representations; see the review \cite{feng2013recent}. However, spectrogram-based analysis methods are commonly used. The spectrogram provides an interpretable form, where both informative (SOI) and non-informative (noise) frequency bands can be identified. The TFR provides an informative data representation that could be visually inspected or used for further automatic processing. Spectrogram analysis for identification of informative frequency bands was proposed by Antoni in \cite{antoni2006spectral}, where spectral kurtosis (in rotating machine diagnostics) was used to identify the frequency content related to SOI. The authors in \cite{Obuchowski2014389} proposed the method for the automatic enhancement of a time-frequency map. It provides a much clearer time-frequency map, and its further application to local damage detection in rotating machines appears to be very effective.
However, the idea of automatic selection of informative spectral content appeared to be very useful, and many researchers contributed to this field. There are other selectors to detect an informative frequency, such as the Gini index \cite{miao2017improvement_GINI}, smoothness index \cite{wang2018some}, negentropy \cite{antoni2016info}, kurtogram \cite{Xiang2015}, infogram \cite{antoni2016info}, IFBI $\alpha$-gram \cite{schmidt2020methodology} or harsogram \cite{zhao2016detection}. In \cite{Obuchowski2014138}, a set of statistics such as Jarque-Bera, Kolmogorov-Smirnov, Cramer-von Mises, Anderson-Darling, quantile-quantile plot, and the method based on the local maxima approach was applied to verify their IFB selection ability. The selection of an appropriate band has been widely discussed in \cite{schmidt2020methodology}.
The detection of the frequency band that contains the SOI should be considered as a pre-processing step in damage detection. It provides "less" noisy time series with clear visibility of cyclic and impulsive SOI, however, the diagnostic information is related to the period between impulses.
The periodicity of the impulses can be detected using the envelope spectrum \cite{randall2011rolling}, which requires the mentioned prefiltering of the signal before demodulation. 
On the other hand, another powerful approach to detect local damages is based on the cyclostationary properties of SOI \cite{antoni2009cyclostationarity, ni2021novel,luo2020cyclic}. 
It should be mentioned here that both envelope analysis or cyclostationary analysis require approximately a constant period between impulses. A real challenge is the diagnosis of bearings with a time-varying speed that results in a time-varying period between impulses, which was reported in \cite{ Sun2021,Kang20157749,Tang20205535}.

In this study, we focus on the constant-speed case and the automation of spectrogram analysis for informative band selection. Since the magnitude part of the spectrogram matrix contains only non-negative elements, the NMF can be applied to decompose the spectrogram into two matrices containing frequency features and temporal features of the input signal. NMF is a commonly used tool in machine learning, blind source separation, or text mining \cite{gillis2014and,cichocki2009nonnegative,fevotte2009nonnegative}. Recently, NMF and its extension NTF have been applied to local fault detection as a source separation technique \cite{gabor2023non}. NMF as an OFB selector was proposed in \cite{wodecki2019novel}. However, in the case of NMF as OFB selector, the obtained filters detect the desired frequency related to damage, but additionally, the filters encompass frequency bands related to noise. To overcome this issue, in this study, the application of the NMU is proposed as an OFB selector. NMU provides a more sparse and more localized part-based decomposition than NMF, resulting in a more selective filter. NMU was previously useful for blind hyperspectral unmixing \cite{casalino2017sequential}, source separation of PET images \cite{kopriva2016single} or climate data analysis \cite{tepper2017nonnegative}. 
Recently, Zhang et al. \cite{Zhang20231} proposed the structured joint sparse orthogonal non-negative matrix factorization for fault detection.  However, to the best of our knowledge, this is the first work that uses NMU to find filter characteristics in a fault detection problem.

The paper is organized as follows. In Section \ref{sec:introduction} the problem formulation and related works are surveyed. The basic theory and NMU-based diagnosis of local damage are described in Section \ref{sec:nmu}. The description of experiments and the two analyzed signals are presented in Section \ref{sec:sexp_setup}. Section \ref{sec:results} shows the obtained results for three compared local damage detection methods: NMU, NMF, and SK. The study is summarized with possible further directions in Section \ref{sec:conclusions}.

\section{NMU-based diagnosis}
\label{sec:nmu}
The approach presented here focuses on obtaining an effective filter, which can be used to detect SOI in the input signal. The input signal can be considered as the mixture of impulsive and periodic SOI and disturbing zero-mean Gaussian noise, which mainly takes the form of colored noise. Using the STFT, the input signal $x(t)$ is transformed into a spectrogram. The magnitude part of the spectrogram is in fact a non-negative matrix $\bS \in \Real_+^{I \times K}$, which is a non-negative linear combination of sources (SOI and noise). The spectrogram $\bS$ can be decomposed into two separate matrices $\bW$ and $\bV$, where $\bW$ is a dictionary matrix containing the frequency components (filters/selectors) and $\bV$ is an activation matrix containing the time information of the signal. The decomposition of the spectrogram can be summarized as:
\be
\label{eq_5} \bS & \cong & \bW \bV, 
\ee
which takes the form of the basic NMF model and was investigated in the area of local damage detection in bearings \cite{wodecki2019novel,wodecki2019impulsive}.

As noted in \cite{wodecki2019novel}, the column vectors of $\bW$ matrix can be used as OFB selectors and, ideally with rank $J=2$ the NMF should give two frequency components, which correspond to the SOI and noise of the input signal. However, in real scenarios, this is not that simple. The real noise takes a more complex structure, and the additional non-linear effects, such as reverberations of the signal, complicate this issue, and the bands of the frequency components penetrate each other, which worsens the detection of SOI. However, as practice shows \cite{wodecki2019novel} for the frequency component that corresponds to SOI, the highest amplitude is pointed to the frequency band related to SOI, and the lower amplitudes point to the frequency bands related to noise. To obtain more selective and sparse frequency components, the NMU can be applied to this task, which by its nature provides a more sparse solution than the NMF. 

Using the Euclidean distance as a cost function and considering the following upper bound constraint $\bW\bV \le \bS$ to the NMF model, the NMU model is obtained. The NMU problem can be reformulated as the following minimization problem:
\be \label{nmu_eq} 
\min_{{\bf W} \in \Real^{I \times J}_+, {\bf V} \in \Real^{J \times K}_+} || \bS - \bW\bV ||_F^2, 
\ee 
where $\bW \ge 0, \bV \ge 0$, $\bW\bV \le \bS$ and $1 \le J < min(I,K).$ The intuitive explanation of obtaining a sparser decomposition of NMU compared to NMF is based on the constraint $\bS \le \bW\bV$, which forces more elements in the matrices to be closer to zero, and the results in the experimental section confirm that. The detailed mathematical sparsity analysis of NMU can be found in \cite{gillis2010using}.

The NMU problem is solved using the Lagrangian relaxation-based algorithm \cite{gillis2010using} and can be applied directly (G-NMU) to the $J$-rank decomposition or can be computed recursively (R-NMU). However, in our experiments, better results were obtained using G-NMU and, as noted in the experimental section in \cite{gillis2010using} the G-NMU obtained a more sparse solution than R-NMU, so this algorithm will be used in our experiments. 

\section{Experimental setup}
\label{sec:sexp_setup}

\begin{figure}[h!]
    \centering
    \includegraphics[scale=0.15]{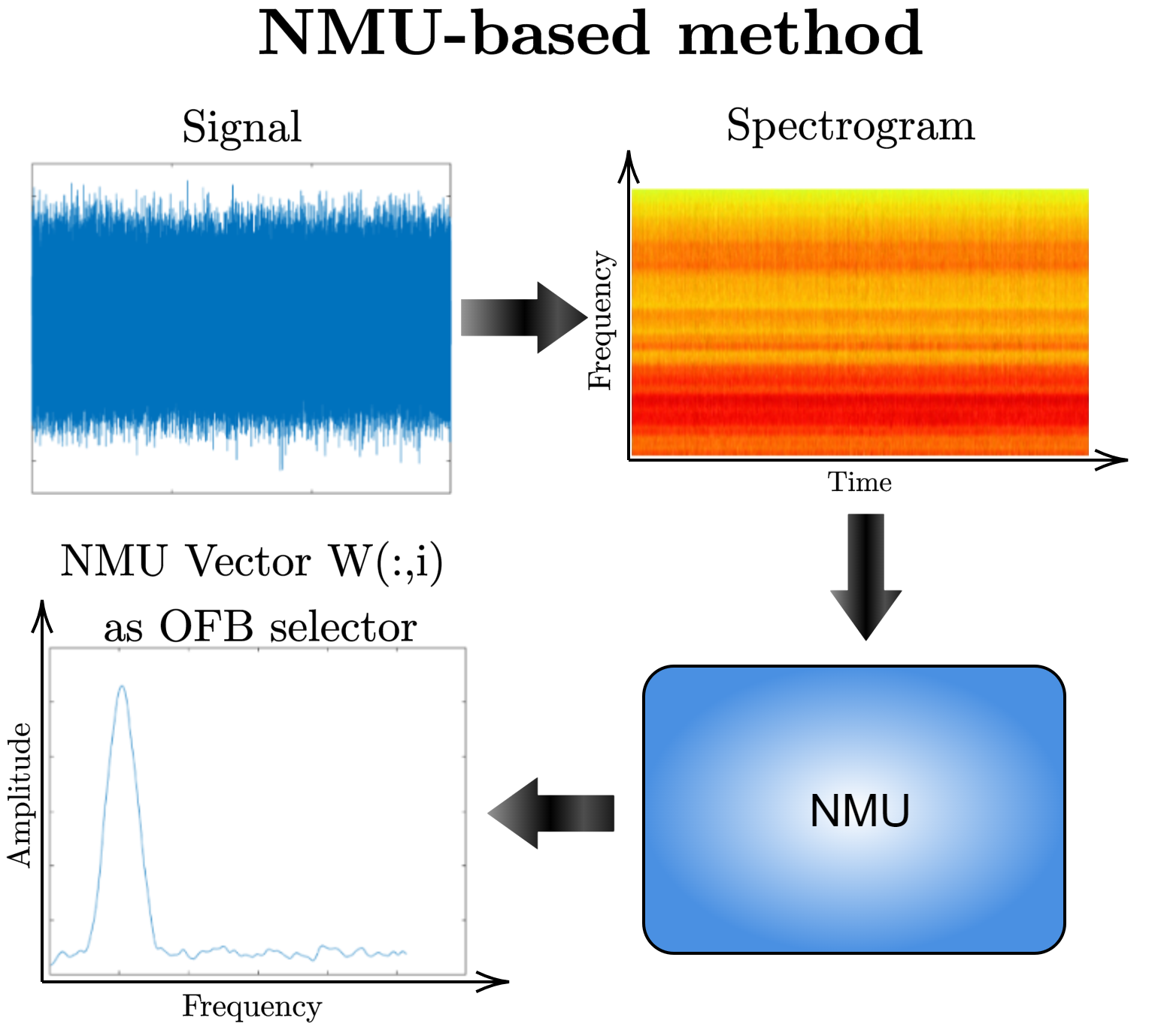}
    \caption{Flowchart of proposed method.}
    \label{fig:flowchart_nmu}
\end{figure}

The input to the experiments is the signal, which is a mixture of SOI and noisy components. The input signal is transformed into a spectrogram and then decomposed into $\bW$ and $\bV$ matrices using the NMU method. The matrix $\bW$ contains $J$ optimal frequency band (OFB) selectors, which are used to filter the original signal. The filtered signals are evaluated using kurtosis, which is a popular statistic used in diagnostics. The best filters are selected based on the highest kurtosis value of the filtered signals. The flowchart of the entire procedure is presented in Figure \ref{fig:flowchart_nmu}. 

The NMU method is compared with two well-known OFB selectors: NMF with multiplicative updates \cite{wodecki2019novel,lee1999learning} and spectral kurtosis \cite{antoni2006spectral}. Both NMF and NMU were evaluated for ranks in the range of 2-15. Because NMF and NMU are very sensitive to initialization, 100 trials with random initialization were performed for each rank, resulting in filter banks of size $13 \times 100$ ($\#$ranks $\times$ 100 trials). For each rank, the average kurtosis is calculated (among 100 trials) and the highest average kurtosis is taken into analysis (Table \ref{tab:if_selectors}). At the end from the filter bank for the given rank with the highest average kurtosis, the filter whose kurtosis is equal to the average kurtosis is selected (Figure \ref{fig:best_gauss}, Figure \ref{fig:best_gauss_acu}). For SK, only one experiment is performed, because SK does not depend on initial conditions and always returns the same result. Spectrograms of the tested signals were computed using the Hamming window of length equal to 128, the overlap was 100 samples, and 512 DFT points were used.

In this work, two real signals with Gaussian noise are evaluated. The first signal is the vibration signal from the test rig with the fault bearing. The signal was recorded using the Kistler 8702B500 accelerometer. The fault frequency is equal to 91.5 Hz, and the informative frequency band is around 20 kHz. This signal is presented in Figure \ref{fig:vibration}. The signal lasts 1 second and the sampling frequency is 50 kHz.

The second example used for validation is the sound of an idler (an element of a belt conveyor system used to transport bulk materials in the mining industry). The signal was recorded with a smartphone. The duration of the acquired signal is 1 second, with a sampling frequency of 48 kHz. The expected fault frequency here is equal to 5.5 Hz, and the informative frequency band is around 2.5 kHz (center frequency). This signal is shown in Figure \ref{fig:acoustic}.

It is worth mentioning that in both cases the cyclic periodic component is hardly seen in both time- and time-frequency domains, thus requires pre-filtering

\section{Results}
\label{sec:results}
The factorized spectrogram is presented in the form of the filter characteristics and the extracted SOI. The effectiveness of the analyzed method can be visually or quantitatively evaluated (the kurtosis of the SOI). The visual criterion for evaluating the filter characteristic aims to select the right informative frequency band and to find the cyclicity and impulsiveness of the SOI after filtering.

\subsection{Vibration signal}
\begin{figure}[h!]
     \centering
     
     \begin{subfigure}[b]{\linewidth}
        \centering
        \includegraphics[scale=0.32]{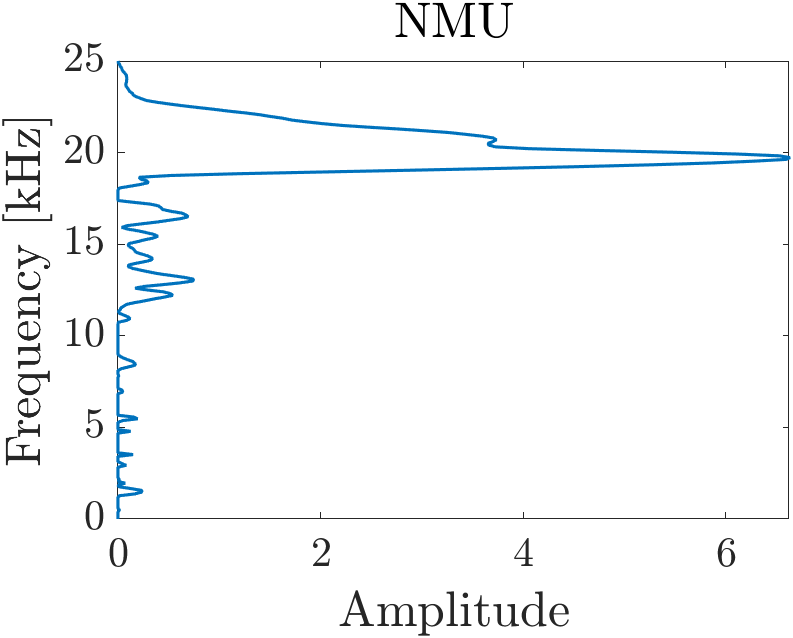}%
        \hfill
        \includegraphics[scale=0.32]{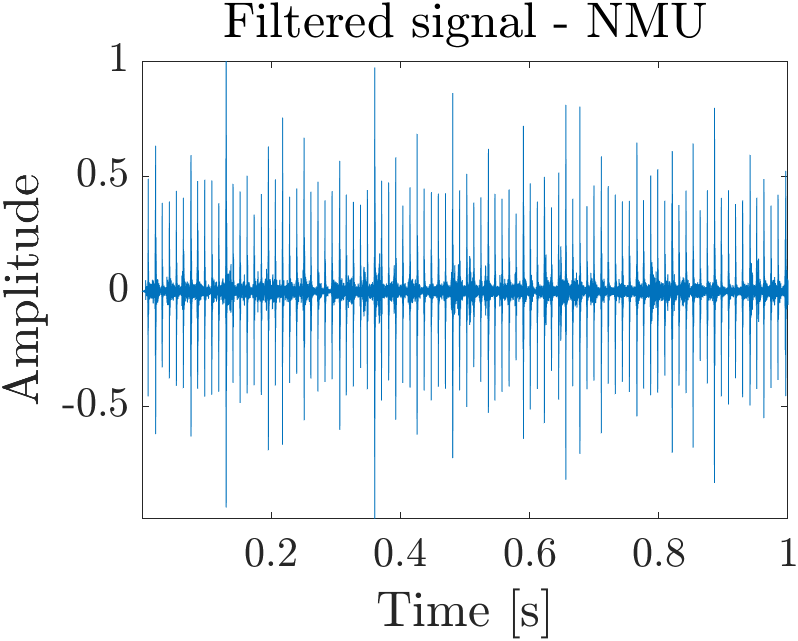}
        \caption{}
    \end{subfigure}
    
    \begin{subfigure}[b]{\linewidth}
        \centering
        \includegraphics[scale=0.32]{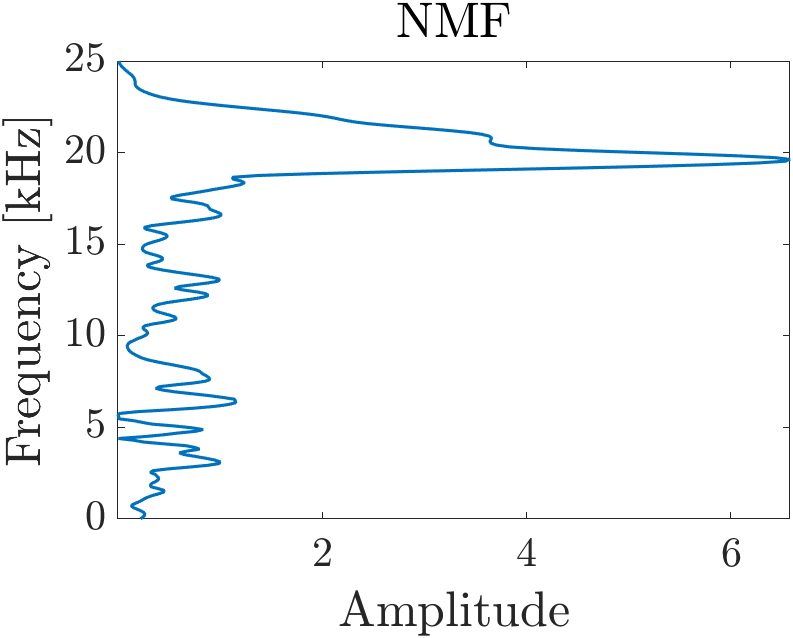}%
        \hfill
        \includegraphics[scale=0.32]{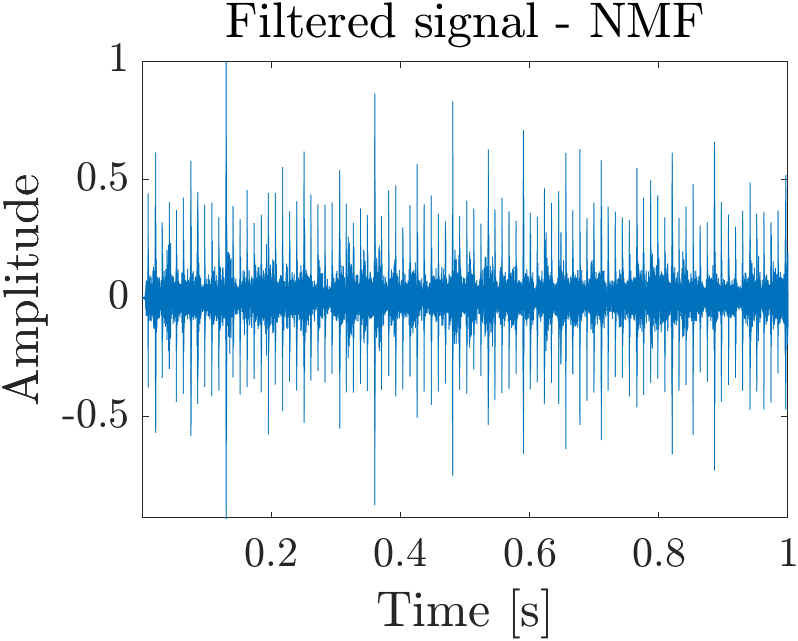}
        \caption{}
    \end{subfigure}
    
    \begin{subfigure}[b]{\linewidth}
        \centering
        \includegraphics[scale=0.32]{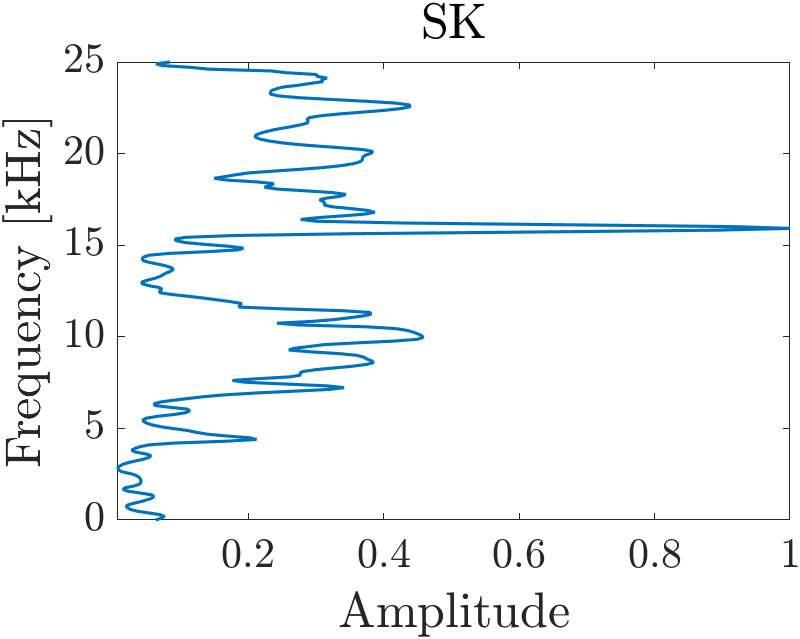}%
        \hfill
        \includegraphics[scale=0.32]{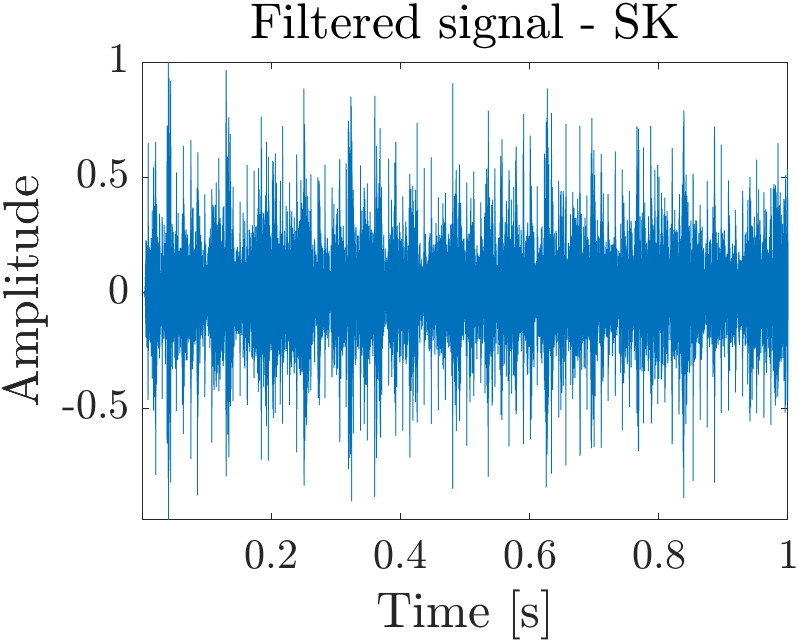}
        \caption{}
    \end{subfigure}
    \caption{Obtained filters and filtered original vibration signal with the following methods: (a) NMU, (b) NMF, (c) SK. The kurtosis values of the filtered signals for the compared methods is presented in Table \ref{tab:if_selectors} (a).}
    \label{fig:best_gauss}
\end{figure}

We can observe in Table \ref{tab:if_selectors} (a) that the kurtosis of the filtered input signal averaged over 100 trials is greater for NMU than for NMF. Compared to the classical SK approach, the average kurtosis obtained by NMU is about five times higher than for SK. To visually evaluate results, the obtained filter characteristics and SOIs are presented in Figure \ref{fig:best_gauss} for NMU, NMF, and SK. As can be noted, both NMF and NMU provide better results than the classical SK approach. The filtered signals are clearly impulsive, cyclic, and contain a much lower level of background noise. It is clearly visible that the SK-based filtered signal is more noisy as the SK-based filter covers many frequency bins and does not select an informative band properly. The NMU method seems to be better than NMF. The exemplary filter characteristics obtained by NMU can be characterized by lower amplitudes of the non-informative frequency bands (below 20 kHz), which are related to the noise and are close to or equal to zero. However, for the NMF case, the non-informative frequency bands are not as small as in NMU, which results in a higher standard deviation of noise in the filtered signal. 

\subsection{Acoustic signal}
\begin{figure}[h!]
     \centering
     
     \begin{subfigure}[b]{\linewidth}
        \centering
        \includegraphics[scale=0.32]{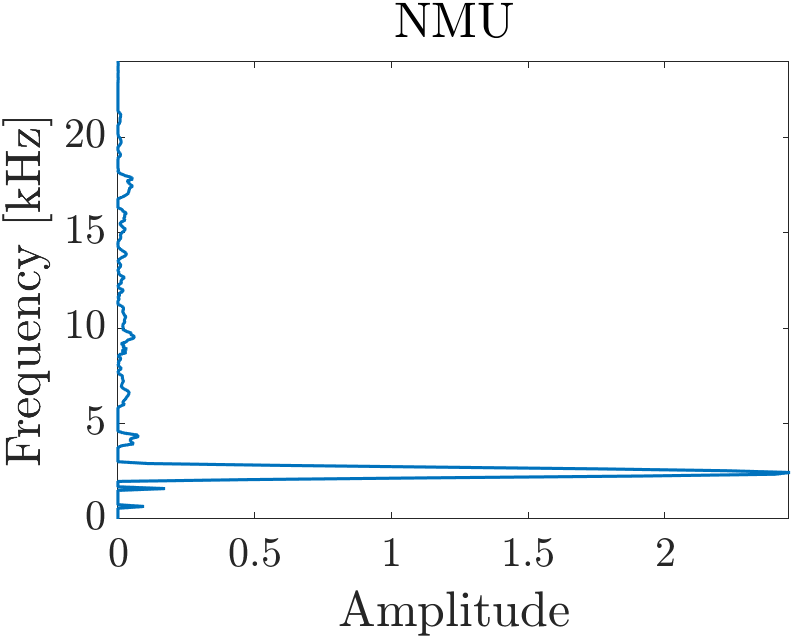}%
        \hfill
        \includegraphics[scale=0.32]{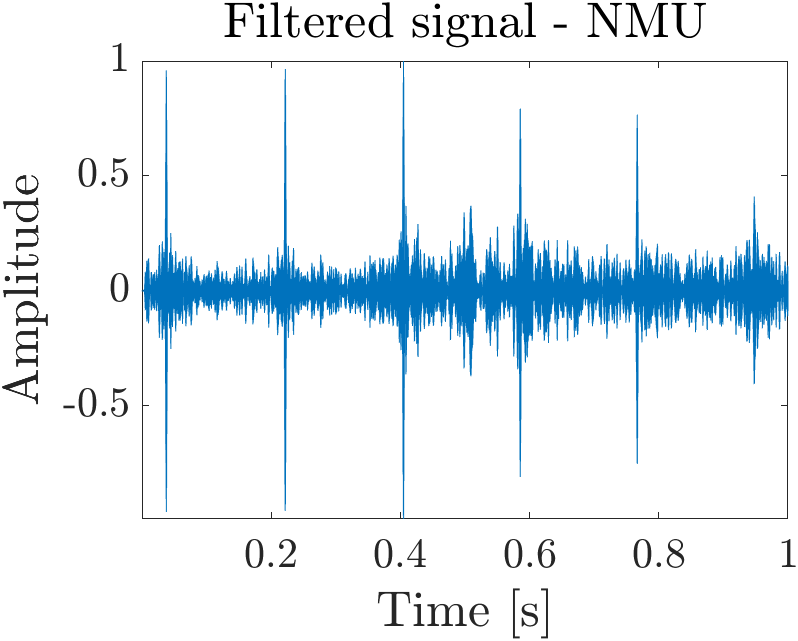}
        \caption{}
    \end{subfigure}
    
    \begin{subfigure}[b]{\linewidth}
        \centering
        \includegraphics[scale=0.32]{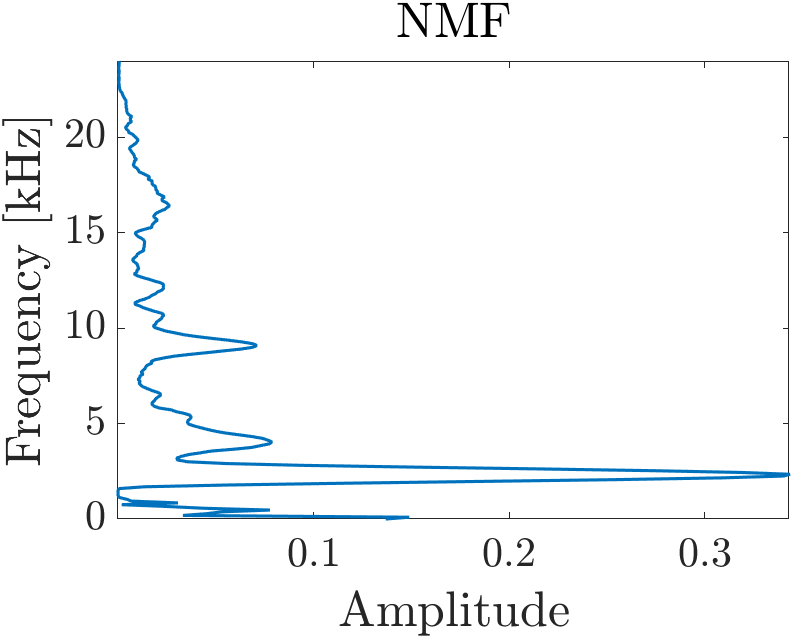}%
        \hfill
        \includegraphics[scale=0.32]{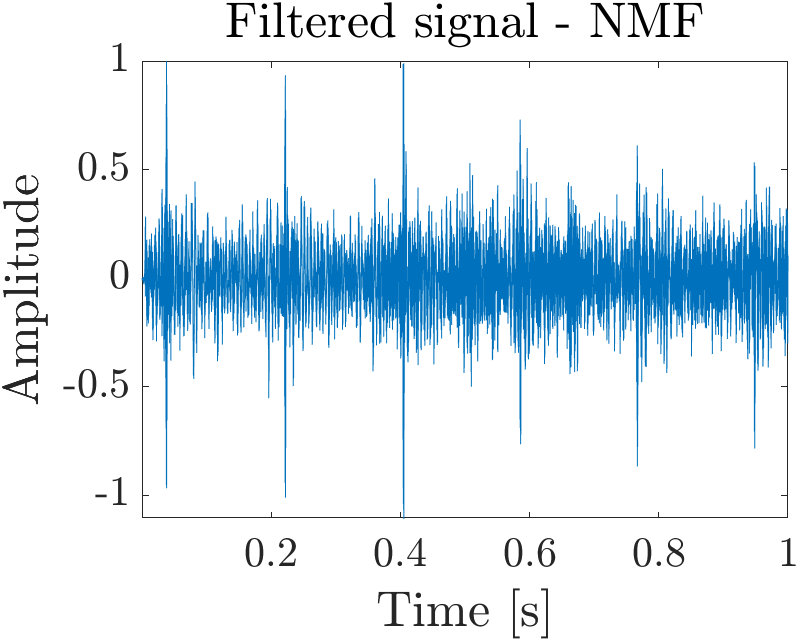}
        \caption{}
    \end{subfigure}
    
    \begin{subfigure}[b]{\linewidth}
        \centering
        \includegraphics[scale=0.32]{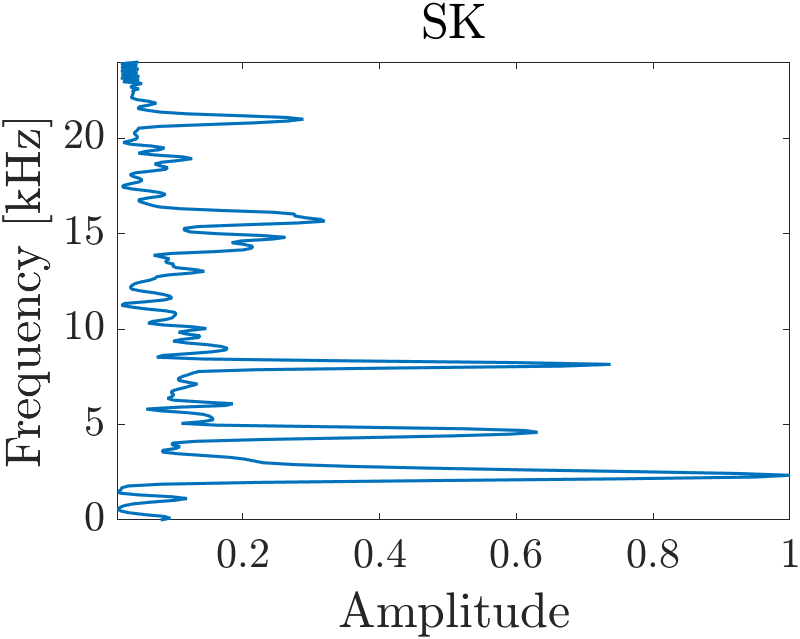}%
        \hfill
        \includegraphics[scale=0.32]{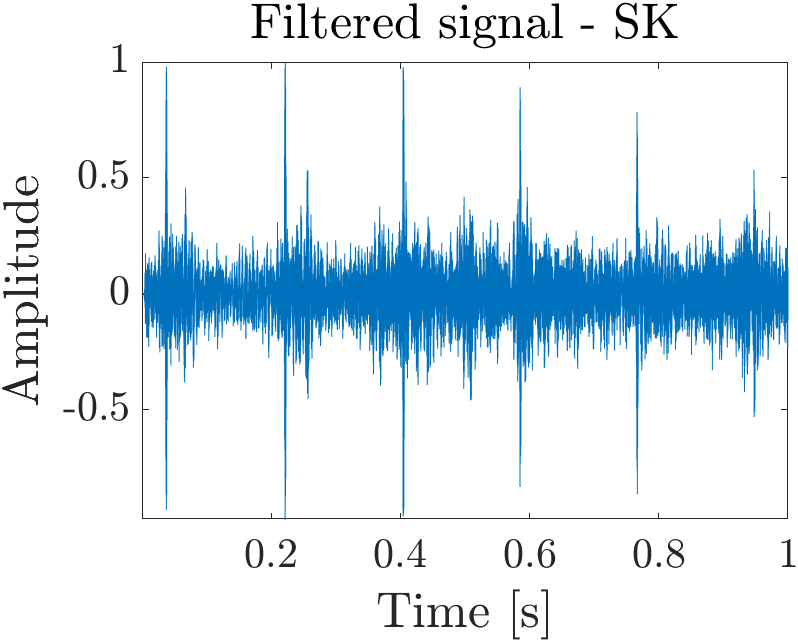}
        \caption{}
    \end{subfigure}
    \caption{Obtained filters and filtered original acoustic signal with the following methods: (a) NMU, (b) NMF, (c) SK. The kurtosis values of the filtered signals for the compared methods is presented in Table \ref{tab:if_selectors} (b).}
    \label{fig:best_gauss_acu}
\end{figure}

The acoustic signal from an idler is more complicated and noisy than the vibration signal. The results for the acoustic signals are presented in Table \ref{tab:if_selectors} (b). Similarly, as in the case of the vibration signal, the average kurtosis is the largest for the NMU. What is interesting is that the SK achieves a larger kurtosis than the average kurtosis of the NMF approach. From this, it can be concluded that, if there are fever cyclic impulses (5.5 Hz), the filter characteristics obtained from NMF have a lower quality. However, this is not the case for NMU. The exemplary filter characteristics obtained with NMU, NMF, and SK for the acoustic signal are shown in Figure \ref{fig:best_gauss_acu}. As can be seen, all the filtered signals are much more impulsive, and the impulses are cyclic. The difference is not as significant in the SOI shape. However, the obtained filter characteristics differ from each other. The most clear filter shape is obtained using the proposed NMU which correctly detects an informative frequency band, and due to the more sparse solution compared to the NMF, the amplitudes of the other frequency bands are close to zero. The classical SK method provides a complex spectral structure of the filter, highlighting many bands. As the more complex characteristic of the filter provides the same results as NMU or NMF, one may conclude that these bins are very low-energy components, and thus may be neglected. A simple filter that provides the same information is better. To conclude, we can say that again, NMU provides the best results.

\begin{table}[h!]
\caption{Kurtosis of the filtered signal obtained with the compared OFB selectors (average for NMF and NMU) for vibration signal (a) and acoustic signal (b).}
\centering
\begin{tabular}{@{}lcllc@{}}
\multicolumn{2}{c}{(a)}               &  & \multicolumn{2}{c}{(b)}            \\ \cmidrule(r){1-2} \cmidrule(l){4-5} 
\textbf{Method}   & \textbf{Kurtosis} &  & \textbf{Method}   & \textbf{Kurtosis} \\ \cmidrule(r){1-2} \cmidrule(l){4-5} 
NMU       & $25.81 $           &  & NMU     &   $17.87 $    \\
NMF \cite{wodecki2019novel} & $11.45  $  &  & NMF \cite{wodecki2019novel}     &    $6.17 $\\
SK \cite{antoni2006spectral} & 5.478             &  & SK \cite{antoni2006spectral} &   11.34    \\ \cmidrule(r){1-2} \cmidrule(l){4-5} 
\end{tabular}
\label{tab:if_selectors}
\end{table}

\section{Conclusions}\label{sec:conclusions}
This study contributes to the area of local damage detection in rolling element bearings. We proposed to factorize the spectrogram matrix using the non-negative matrix under-approximation, and then using the frequency-based features of matrix $\bW$ as an optimal filter. The results presented here emphasize the importance of sparsity to obtain a more selective filter, which mostly covers an informative part of the signals, neglecting the frequency bands related to noise. This translates into a lower amplitude of the noise component in the filtered signal, resulting in a higher kurtosis value. Due to the under-approximation constraint in NMU, sparsity is naturally obtained. The proposed approach was compared with previously used methods, such as NMF and SK. NMU provides a more selective filter than NMF and SK. The current study was limited to the signal with Gaussian noise only. In the future, the proposed method should also be applied to signals with non-Gaussian noise. Furthermore, as done in \cite{gillis2013sparse}, future works should investigate the extension of the proposed approach by enhancing the sparsity properties of NMU incorporating additional sparsity constraints, extending the proposed idea to analyze faults in electric motors, and evaluating on more challenging data from different conditions.

\section{Statements}
The work is supported by the National Center of Science under Sheng2 project No. UMO-2021/40/Q/ST8/00024 "NonGauMech - New methods of processing non-stationary signals (identification, segmentation, extraction, modeling) with non-Gaussian characteristics for the purpose of monitoring complex mechanical structures".

\bibliographystyle{IEEEtran}
\bibliography{mybib}

\end{document}